\documentclass[final,3p,times,twocolumn]{elsarticle}

\usepackage{graphicx}
\usepackage{dcolumn}
\usepackage{bm}
\usepackage{multibib}
\usepackage{adjustbox}
\usepackage{amsmath, amssymb, amsthm, thm-restate, bbold}
\usepackage{braket, physics, enumitem, relsize}
\usepackage{makecell, graphicx, subfiles, multirow, wrapfig}
\usepackage[font=scriptsize]{subcaption}
\usepackage[font=scriptsize, justification=raggedright]{caption}
\usepackage{subcaption}
\usepackage[toc,page]{appendix}
\usepackage[colorlinks, allcolors=black]{hyperref}
\usepackage[capitalize, nameinlink]{cleveref}
\usepackage{xcolor, tabularx}
\setlist[itemize]{leftmargin=10pt}
\usepackage[utf8]{inputenc}
\usepackage{float}
\usepackage{tikz}
\usepackage{float} 
\usepackage{ulem}
\usepackage{flushend}
\usepackage{rsfso}
\usepackage{mathrsfs}
\usepackage{mathalfa}
\usepackage{frcursive}
\usepackage{comment}
\usepackage{dblfloatfix} 
\usepackage{cuted} 

\newenvironment{frcseries}{\fontfamily{frc}\selectfont}{}

\usetikzlibrary{quantikz}
\usepackage{xcolor}
\usepackage{hyperref}
\hypersetup{
    colorlinks=true,
    citecolor=blue,
    linkcolor=blue,
    filecolor=magenta,      
    urlcolor=cyan,
    pdftitle={Overleaf Example},
    pdfpagemode=FullScreen,
    }

\mathcode`l="8000
\begingroup
\makeatletter
\lccode`\~=`\l
\DeclareMathSymbol{\lsb@l}{\mathalpha}{letters}{`l}
\lowercase{\gdef~{\ifnum\the\mathgroup=\m@ne \ell \else \lsb@l \fi}}%
\endgroup

\renewcommand{\epsilon}{\varepsilon}

\newtheorem*{thm*}{Theorem}

\renewcommand{\vec}[1]{\overrightarrow{#1}}

\usepackage{ragged2e}
\makeatletter
\newcounter{savesection}
\newcounter{apdxsection}
\renewcommand\appendix{\par
  \setcounter{savesection}{\value{section}}%
  \setcounter{section}{\value{apdxsection}}%
  \setcounter{subsection}{0}%
  \gdef\thesection{\@Alph\c@section}}
\newcommand\unappendix{\par
  \setcounter{apdxsection}{\value{section}}%
  \setcounter{section}{\value{savesection}}%
  \setcounter{subsection}{0}%
  \gdef\thesection{\@arabic\c@section}}
\makeatother
\usepackage{lineno}

\begin{document}

\begin{frontmatter}



\title{Non-linearity and chaos in the kicked top}


\author[aff1,aff2]{Amit Anand\corref{cor1}} 
\cortext[cor1]{Corresponding author}
\ead{a63anand@uwaterloo.ca}
\author[aff2,aff3]{Robert B. Mann} 
\author[aff1,aff2,aff3,aff4]{Shohini Ghose} 

\affiliation[aff1]{organization={Institute for Quantum Computing, University of Waterloo},
            addressline={200 University Ave W}, 
            city={Waterloo},
            postcode={ N2L 3G1}, 
            state={Ontario},
            country={Canada}}
            
\affiliation[aff2]{organization={Department of Physics and Astronomy, University of Waterloo},
            addressline={}, 
            city={Waterloo},
            postcode={ N2L 3G1}, 
            state={Ontario},
            country={Canada}}
            
\affiliation[aff3]{organization={Perimeter Institute for Theoretical Physics},
            addressline={31 Caroline St N}, 
            city={Waterloo},
            postcode={ N2L 2Y5}, 
            state={Ontario},
            country={Canada}}

\affiliation[aff4]{organization={Department of Physics and Computer Science, Wilfrid Laurier University},
            addressline={75 University Ave W}, 
            city={Waterloo},
            postcode={ N2L 3C5}, 
            state={Ontario},
            country={Canada}}


\begin{abstract}
Classical chaos arises from the inherent non-linearity of dynamical systems. However, quantum maps are linear; therefore, the definition of chaos is not straightforward. To address this, we study a quantum system that exhibits chaotic behavior in its classical limit: the kicked top model, whose classical dynamics are governed by Hamilton's equations on phase space, whereas its quantum dynamics are described by the Schrödinger equation in Hilbert space. 
We explore  the critical degree of non-linearity signifying the onset of chaos in the kicked top by modifying the original Hamiltonian 
so that the non-linearity is  parametrized by a quantity $p$.  We find two distinct behaviors of the modified kicked top depending on the value of $p$. Chaos intensifies as $p$ varies within the range of $1\leq p \leq 2$, whereas it diminishes for $p > 2$, eventually transitioning to a purely regular oscillating system as $p$ tends to infinity. We also comment on the complicated phase space structure for non-chaotic dynamics. Our investigation sheds light on the relationship between non-linearity and chaos in classical systems, offering insights into their dynamic behavior.

\end{abstract}

\begin{keyword}
Classical chaos \sep non-linear dynamics \sep kicked top


\end{keyword}

\end{frontmatter}




\section{Introduction}\label{section:introduction}
The discovery of classical chaos marked a significant milestone that has captured the interest of both physicists and mathematicians ever since. Over the decades,   scientists have endeavoured to understand the underlying properties  that are responsible for the chaotic dynamics of a system. This led to  investigations of many non-linear systems, and an emerging understanding that breaking integrability in such systems constitutes a minimum criterion for exhibiting chaotic behavior \cite{kozlov_integrability_1983}. 

One approach to study  non-integrability in a  Hamiltonian system with $N-$degrees of freedom \cite{,Daniel_2017,ziglin_branching_1983}, involves introducing non-integrable perturbations, where $H = H_0(\vec{I}) + \epsilon H_1(\vec{I},\vec{\theta})$ and $\vec{I}$ and $\vec{\theta}$ are action and angle variables of the Hamiltonian $H$. The term $H_0$ governs the motion on the $N-$torus, with frequencies $\omega_i = \partial H_0/\partial I_i$. According to KAM theory \cite{kolmogorov_1954,arnold_proof_1963,moser_1962}, the integrability of the Hamiltonian system breaks down for any $\epsilon>0$,  although most invariant tori do not vanish entirely but rather undergo slight deformations. For a  majority of initial points in phase space,
this results in   quasi-periodic behavior of the resultant trajectories. As the strength of the perturbation increases, the system transitions to chaotic dynamics. The degree of chaos in Hamiltonian systems is determined by the interplay between the strength of the perturbation and the system's non-linearity.
A natural question then arises as to what degree of non-linearity a system must possess in order to exhibit  chaotic dynamics. Our study is motivated by this query,  leading us to explore a classical system that has a well-defined non-trivial quantum counterpart.  

The importance of non-linearity in classical systems poses an initial challenge when extending the study of chaos to quantum systems. Unlike classical dynamics, quantum dynamics are governed by linear Schrödinger equations, resulting in  insensitivity towards the perturbations in the initial state.  Moreover, the integrability of quantum systems is still not fully understood \cite{Mossel_caux_2011}. As a result, attention has turned to systems such as the cat map, the kicked rotor, and others that exhibit chaotic behavior in their classical limits \cite{Ford_Ristow_cat_1991,fishman_rotor_anderson_1982}. 
A study of the dynamical properties of such systems for the same set of parameters when quantized, 
  provides insight into the emergence of chaos (if present)  from quantum dynamics. One such model -- the kicked top -- was introduced by Haake, Kuś, and Scharf, and  has a chaotic classical limit \cite{haake1987}.  The Hamiltonian \begin{equation}\label{eq:KThamiltonian}
H =\hbar \frac{\alpha J_{y}}{\tau} + \hbar \frac{\kappa J_{z}^2 }{2j}\Bigg( \sum_{n=-\infty}^{n=\infty} \delta (t-n\tau) \Bigg)
 \end{equation}
describing the kicked top consists of constant precession by angle $\alpha$, and a periodic non-linear (quadratic) pulse of  strength $\kappa$. The total angular momentum of the system is constant, and the classical limit is attained by taking $\hbar/j \to 0$. 
Multiple theoretical and experimental investigations of this model have been carried out for more than thirty years in an effort to understand   both classical and quantum chaos  \cite{haake1987,Constantoudis_and_Theodorakopoulos_1997,meenu_correspondence-2018,Google}.

Our focus lies in understanding the critical exponent of the non-linear component of the kicked-top Hamiltonian, marking the transition from regular behavior to chaotic dynamics.  We shall restrict ourselves only to classical dynamics, building upon recent research \cite{Poggi_chaos_2021} that explored the dependence of non-linear exponents by replacing $J_z^2$ with $J_z^p$, where $p$  is an arbitrary  positive integer. In the original kicked top, $p=1$ corresponds to regular dynamics and $p=2$ is fully chaotic for large kicking strength ($\kappa)$
\cite{haake1987}. Our interest lies in exploring the behaviour of the kicked top for values of  $p\neq 1,2$. However introducing  non-integer values of $p$ renders the quantum version of the Hamiltonian  non-Hermitian. To address this, we replace $J_z^p$ with $|J_z|^p$, thus preserving Hermiticity  in the quantum case. This modification enables us to investigate the kicked top for non-integer exponents.  

We use the largest Lyapunov exponent to quantify chaos and study its dependence on $p$. We classify the modified kicked top into two distinct regimes: $1\leq p \leq 2$ and $p > 2$, and analyze their respective behaviors. Our findings demonstrate that the system loses integrability for any $p \geq 1$, leading to the emergence of chaos. Furthermore, we observe that the degree of chaos intensifies with increasing non-linearity for $1\leq p \leq 2$. 
As we further increase $p\,( >2)$, we observe a suppression of chaos, with regions of chaotic dynamics confined to increasingly smaller areas of phase space. This phenomenon give rise to further investigation into the underlying dynamics governing the system's behavior. We explain this behaviour by linearizing the stroboscopic map and analyzing it for different strength of non-linearity ($p$).  For $p=1$, our work also relates to a study of different conservative Hamiltonian models of fractal dimensions on atorus and on cylindrical phase space \cite{scott_hamiltonian_2001,Chen_meiss_1989,ashwin_97}. 

The outline of our paper is as follows. In section \ref{section:p-kicked top model}, we introduce the modified kicked top. In section \ref{section:lyapunov exponent}, we discuss the Lyapunov exponent as a quantifier of chaos in classical system. We present our results in \ref{section: numerical investigations}. In section \ref{section:fractal-like structures}, we discuss the complex structure of the phase space for the case of $p=1$. We conclude our work by discussing the results and prospects for future work in \ref{section: discussion}.

\section{Modulus $p-$kicked top model}\label{section:p-kicked top model}
\begin{figure*}[p] 
    \centering
    \begin{subfigure}{0.495\textwidth} 
        \centering
        \includegraphics[width=\linewidth,height=5cm]{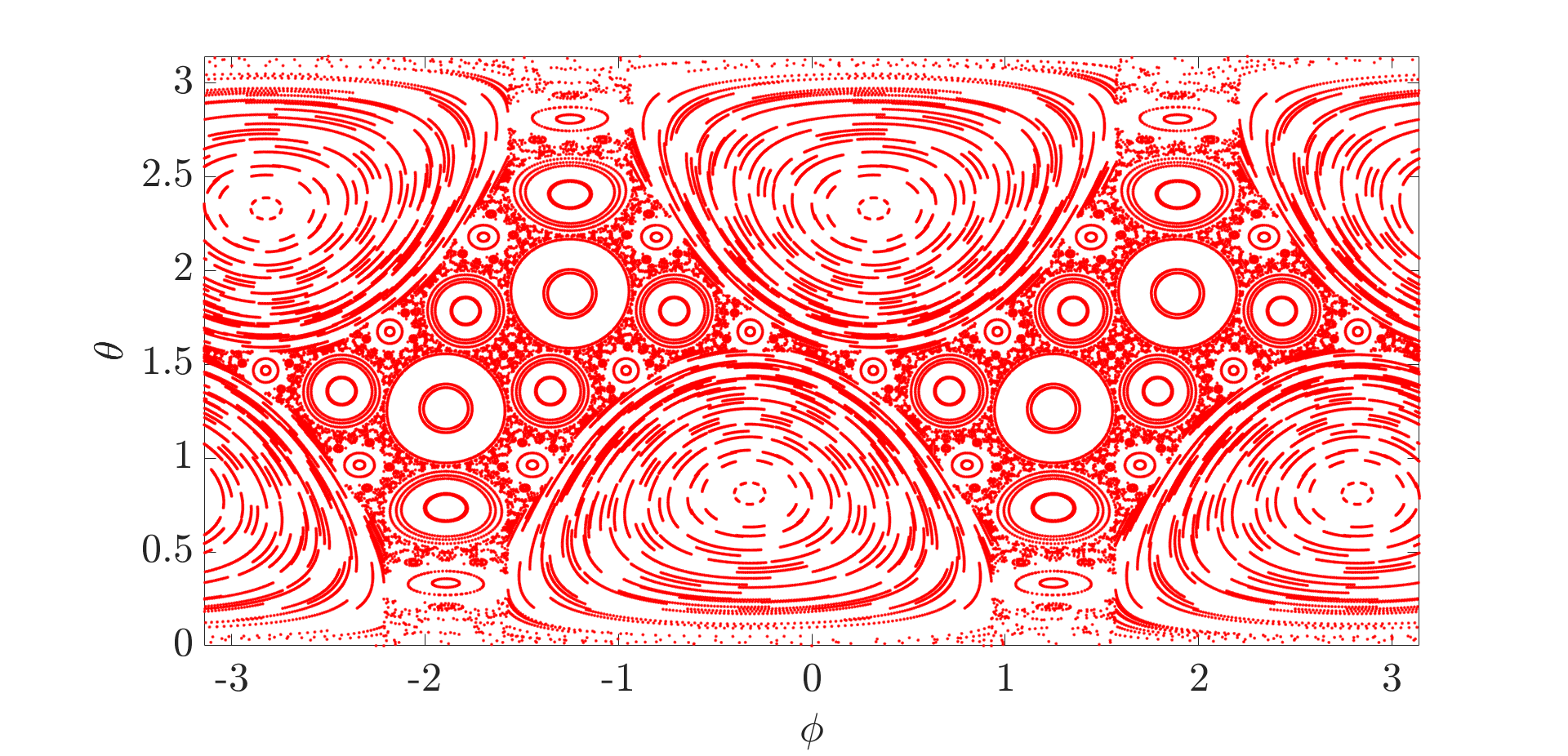}
        \caption{$p=1.0$}\label{fig:p1.0}
    \end{subfigure}
    \hfill
    \begin{subfigure}{0.495\textwidth}
        \centering
        \includegraphics[width=\linewidth,height=5cm]{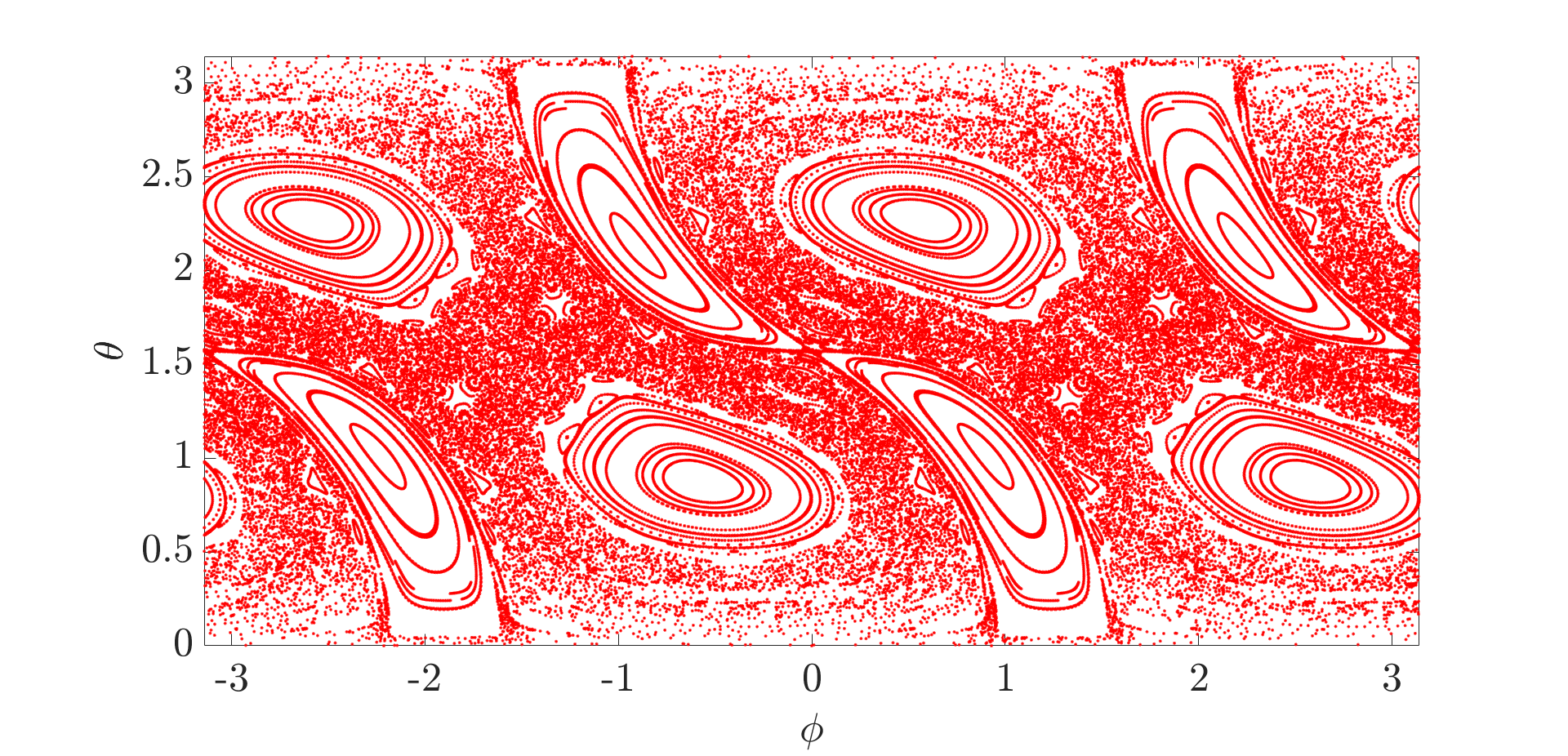}
        \caption{p=1.5}\label{fig:p1.5}
    \end{subfigure}
    
    \vspace{0.5cm} 
    
    \begin{subfigure}{0.495\textwidth}
        \centering
        \includegraphics[width=\linewidth,height=5cm]{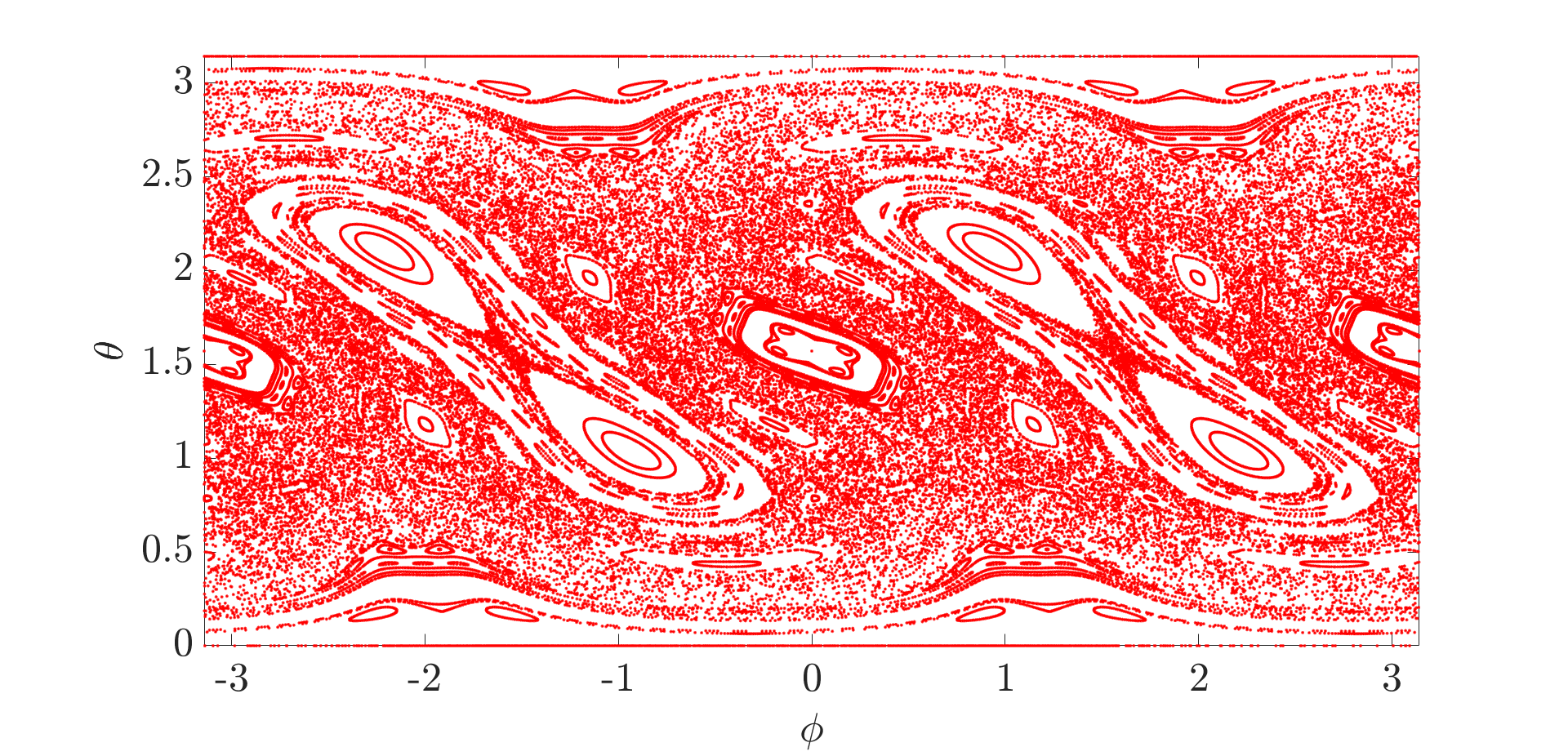}
        \caption{p=2.0}\label{fig:p2.0}
    \end{subfigure}
    \hfill
    \begin{subfigure}{0.495\textwidth}
        \centering
        \includegraphics[width=\linewidth,height=5cm]{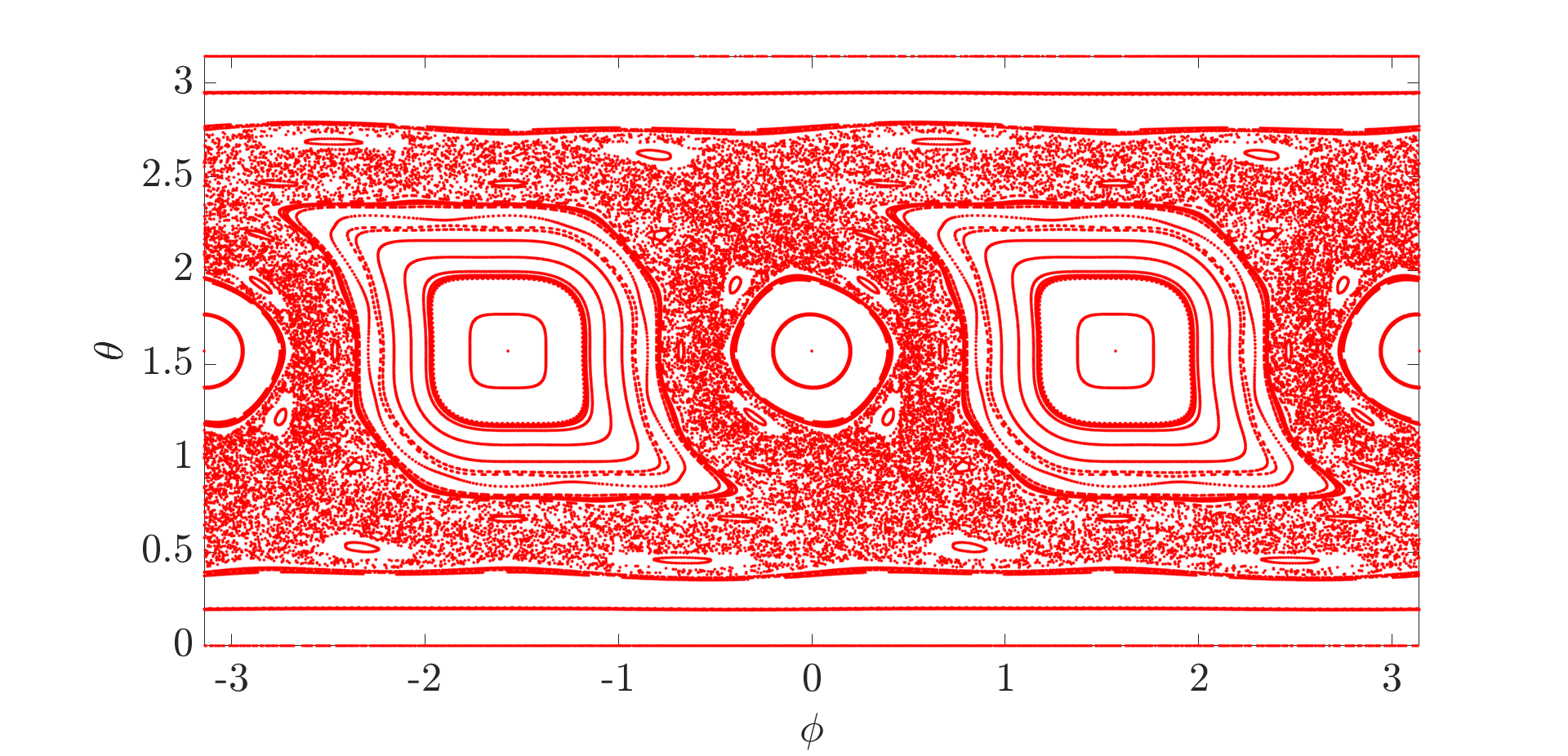}
        \caption{$p=4$}\label{fig:p4}
    \end{subfigure}
    
    \vspace{0.5cm} 
    
    \begin{subfigure}{0.495\textwidth}
        \centering
        \includegraphics[width=\linewidth,height=5cm]{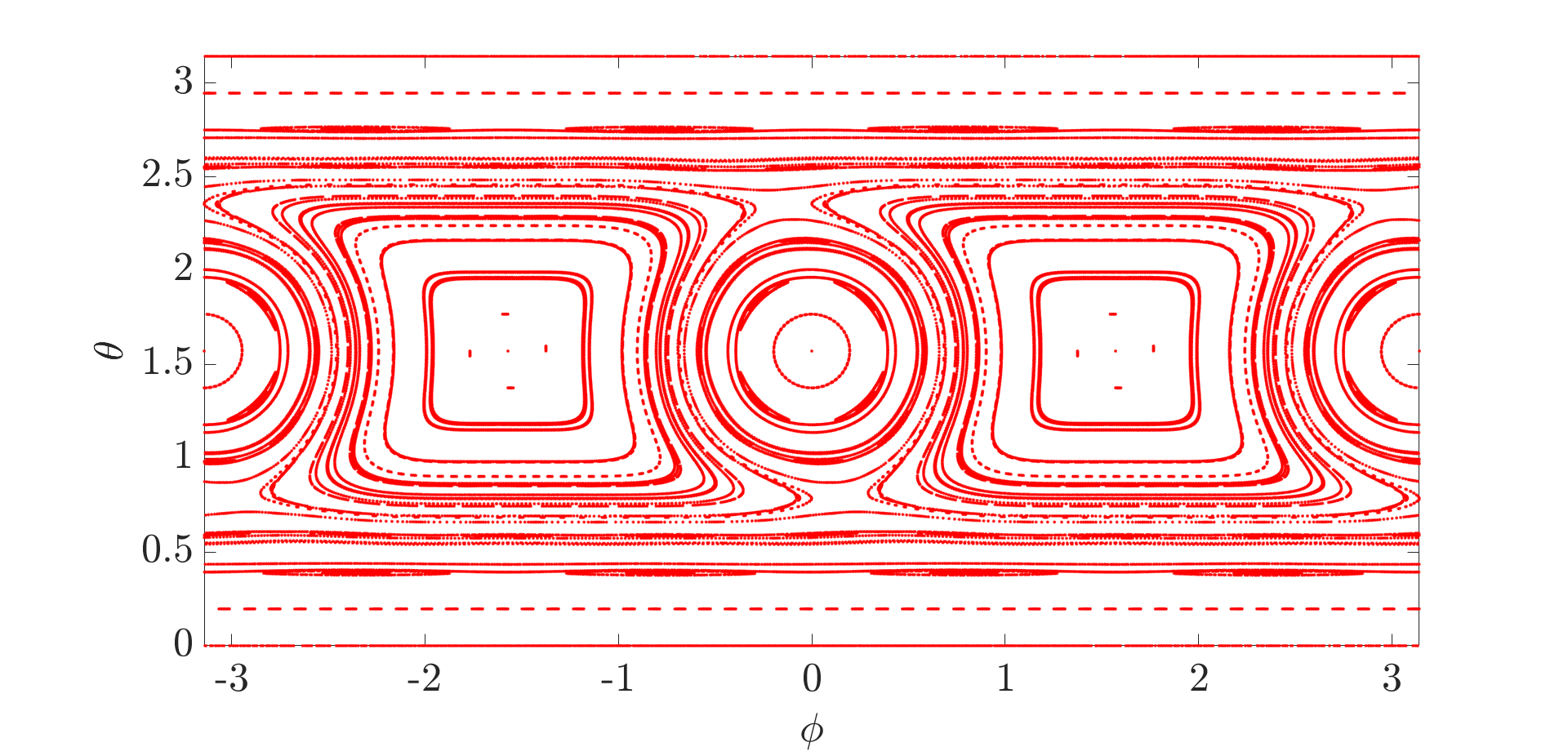}
        \caption{$p=7$}\label{fig:p7}
    \end{subfigure}
    \hfill
    \begin{subfigure}{0.495\textwidth}
        \centering
        \includegraphics[width=\linewidth,height=5cm]{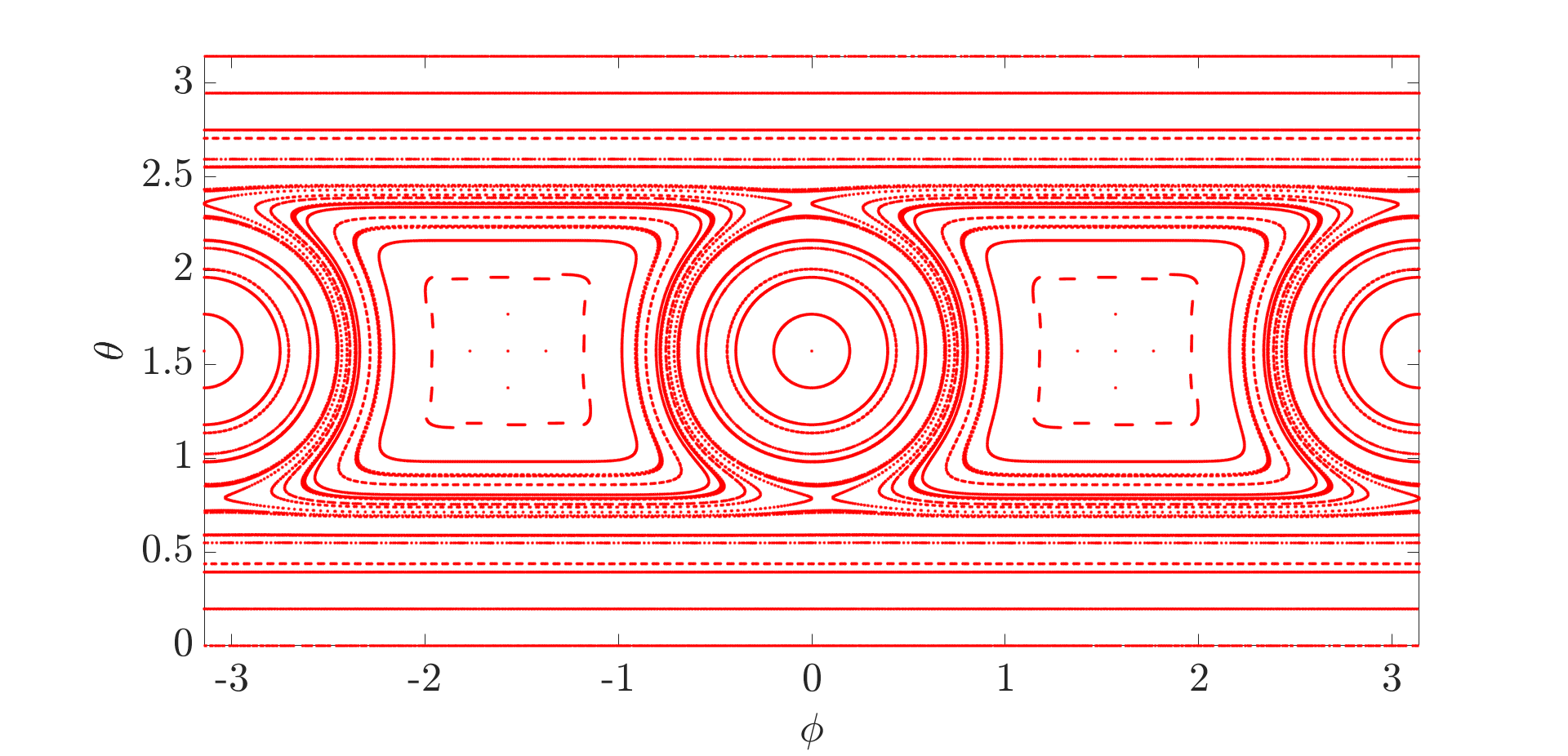}
        \caption{$p=10$}\label{fig:p10}
    \end{subfigure}
    \caption{Stroboscopic map showing the classical time evolution over 400 kicks for 289  initial points uniformly distributed in phase space for $\kappa=2.5$ and $\alpha=\pi/2$  and different values of $p$.}
   \label{fig:classical phase space structure} 
\end{figure*}

We study a modified version of Haake's kicked top \cite{haake1987} that  is a
time-dependent, periodically-driven system   governed by the Hamiltonian
\begin{equation}\label{eq:modulus_hamiltonian}
H =\hbar \frac{\alpha J_{y}}{\tau} + \hbar \frac{\kappa {|J_{z}|}^p}{pj^{p-1}}\Bigg( \sum_{n=-\infty}^{n=\infty} \delta (t-n\tau) \Bigg),
 \end{equation}
where $\{ J_{x}, J_{y}, J_{z}\}$ are the generators of angular momentum, with 
$$
[J_i, J_j] = i\epsilon_{ijk} J_k\; .
$$ 
The Hamiltonian \eqref{eq:modulus_hamiltonian} describes a spin-$j$ object precessing about the $y$-axis together with impulsive state-dependent twists about the $z$-axis whose magnitude is characterized by the chaoticity parameter $\kappa$.  The period between kicks is $\tau$, and $\alpha$ is the amount of $y$-precession within one period. The parameter $p$ determines the non-linearity of the system.

The classical kicked top is obtain by computing the Heisenberg equations for the re-scaled angular momentum generators, $J_i/j$, followed by the limit $j \to \infty$. The resulting stroboscopic map $F$ of the classical coordinates of $\textbf{X} = (X,Y, Z)$ on the unit sphere is given by
\begin{eqnarray}\label{eq:classical_eqn}
\nonumber X_{n+1}&=&\mu_n\cos\bigg(\kappa \zeta_n |\zeta_n|^{p-2} \bigg) - Y_n\sin\bigg(\kappa \zeta_n |\zeta_n|^{p-2}\bigg), \\  
\nonumber Y_{n+1}&=&Y_n\cos\bigg(\kappa \zeta_n |\zeta_n|^{p-2} \bigg) + \mu_n\sin\bigg(\kappa \zeta_n |\zeta_n|^{p-2}\bigg), \\
 Z_{n+1}&=& Z_n \cos(\alpha) - X_n \sin(\alpha),
\end{eqnarray}
 where
\begin{eqnarray}
      \nonumber     \zeta_n = Z_n \cos(\alpha) - X_n \sin(\alpha),\\
      \nonumber     \mu_n = X_n \cos(\alpha) + Z_n \sin(\alpha).
\end{eqnarray}

The derivation of Eq. \eqref{eq:classical_eqn} is given in appendix \ref{Appendix: Classical equation of motion}. The classical dynamics takes place on the unit sphere, and so can be  parameterized using polar and azimuthal angles 
$(\theta,\phi)$. For any initial state $(\theta_0,\phi_0)$, the final state of the system is determined by applying $F^n$ on the initial state, where $n$ is the number of kicks applied. For $p=2$, the model reduces to the original well-studied   kicked top \eqref{eq:KThamiltonian}\cite{haake1987}. As with the original kicked top, for $\alpha=\pi/2$, the generalized kicked top  \eqref{eq:modulus_hamiltonian} also has two fixed points and period-4 points, which are obtained by solving the equation $F^n[\boldsymbol{ X_0}] = \boldsymbol{ X_0}$, except for $p=1$. The fixed points lie at $(0,\pm 1, 0)$ and the points on the period-4 orbit are $\boldsymbol{ X_1}(1,0,0)\rightarrow \boldsymbol{ X_2}(0,0,1)\rightarrow \boldsymbol{ X_3}(-1,0,0)\rightarrow \boldsymbol{ X_4}(0,0,-1)$. A detailed discussion of the classical dynamics of the original $p=2$ kicked top can be found in \cite{haake1987,meenu_correspondence-2018}.

\section{Lyapunov exponent}\label{section:lyapunov exponent}

\begin{figure*}[htbp] 
    \centering
    \begin{subfigure}[H]{0.45\textwidth}
        \centering
        \includegraphics[trim={4cm 0 0 0},clip, width=9.5cm,height=5cm]{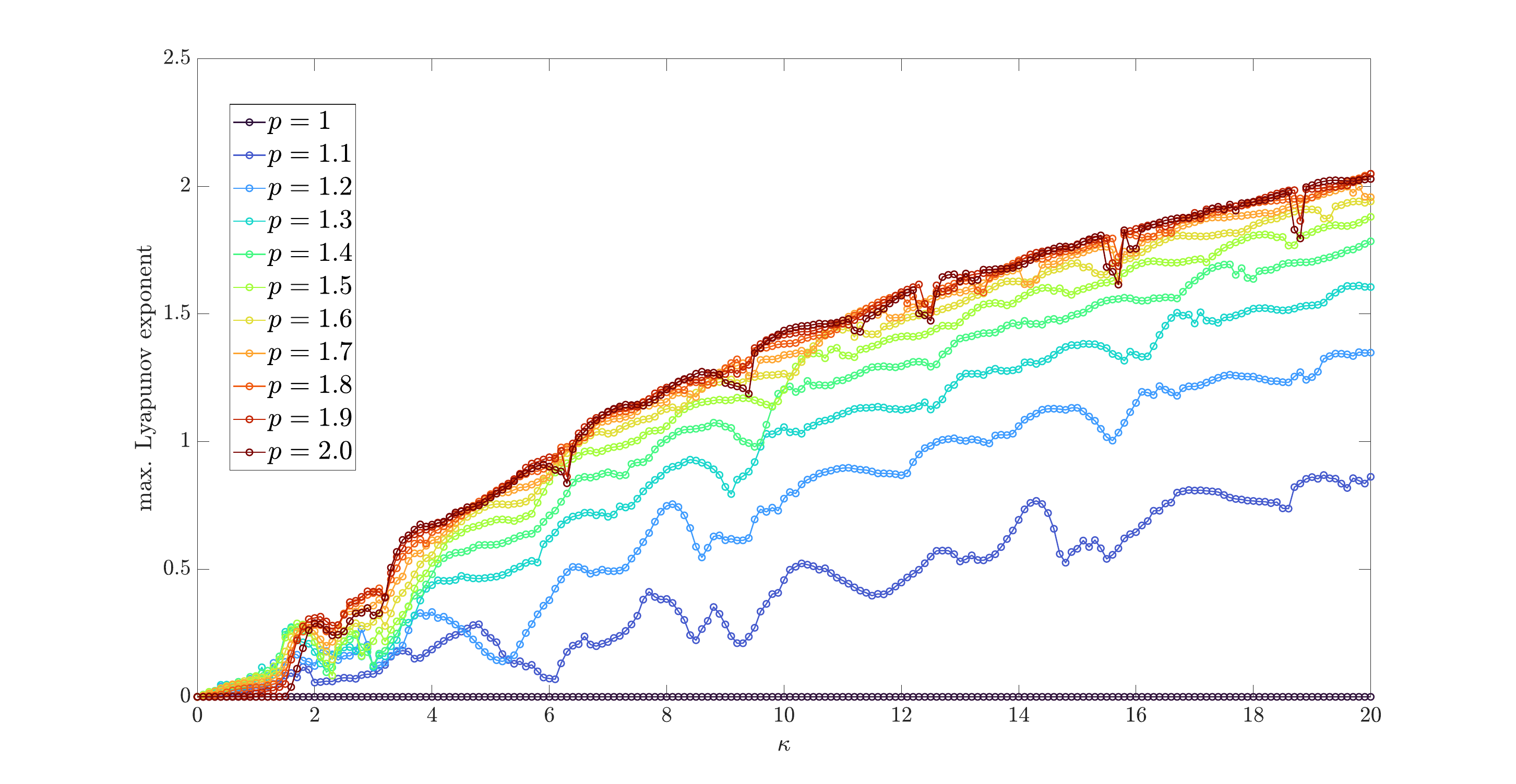}
       \caption{Classical Lyapunov exponent for  $1 \leq p \leq 2$. The $p=1$ curve lies along the $LE=0$ axis.}
       \label{fig:max lyapunov exp. for mod p-spin with 1.0 to 2.0 with 0.1 div model with 10000 kick and 200 div upto k 20}
    \end{subfigure}
    \hfill
    \begin{subfigure}[H]{0.45\textwidth}
        \centering
        \includegraphics[trim={4cm 0 0 0},clip,width=9.5cm,height=5cm]{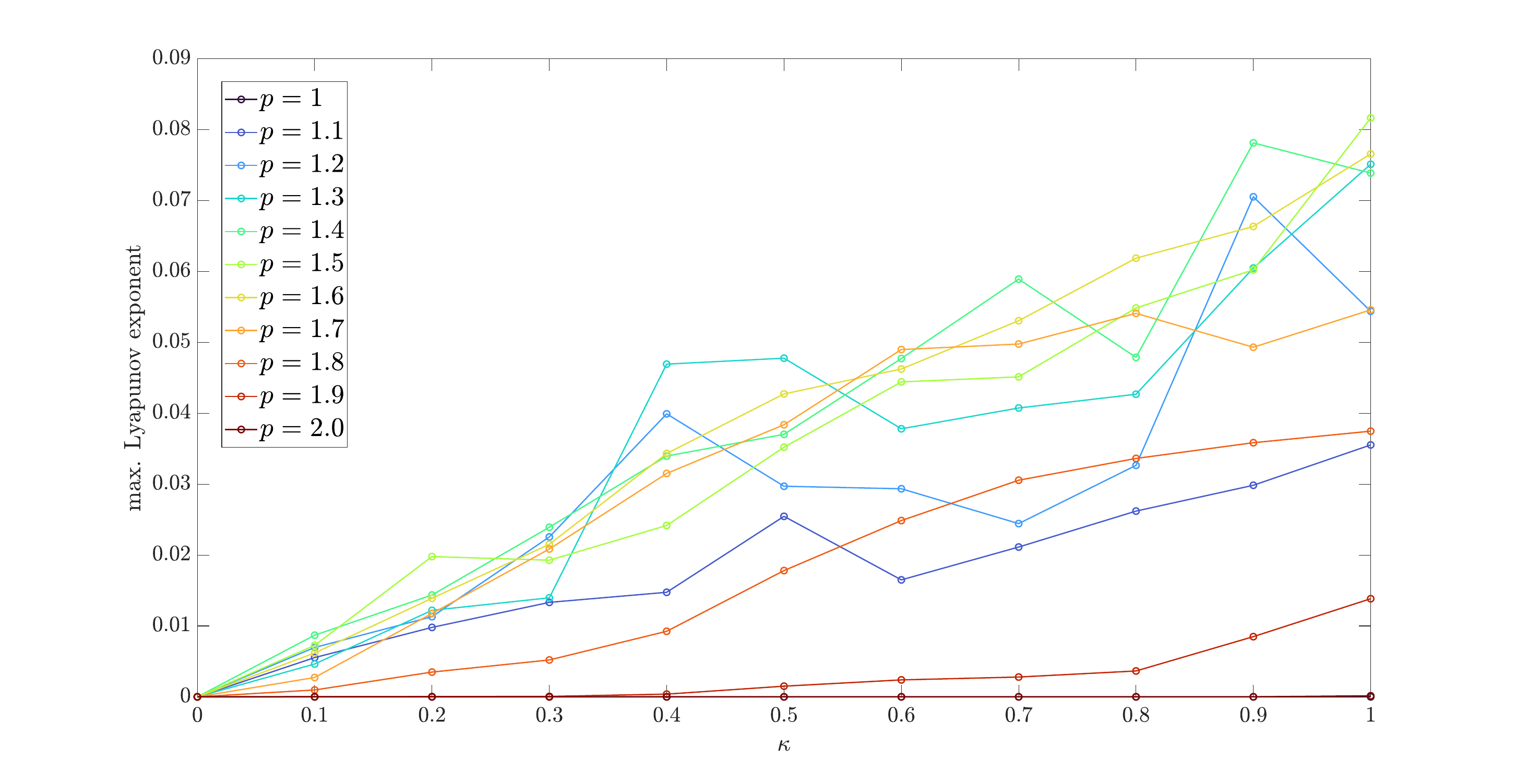}
        \caption{Classical Lyapunov exponent  $1 \leq p \leq 2$. 
        The $p=1$ and $p=2$ curves lie along the $LE=0$ axis }\label{fig:max lyapunov exp. for mod p-spin with 1.0 to 2 with 1 div model with 10000 kick and 200 div upto k 1}
    \end{subfigure}
    \vspace{0.5cm} 
    \\
    \begin{subfigure}[H]{0.45\textwidth}
        \centering
        \includegraphics[trim={4cm 0 0 0},clip,width=9.5cm,height=5cm]{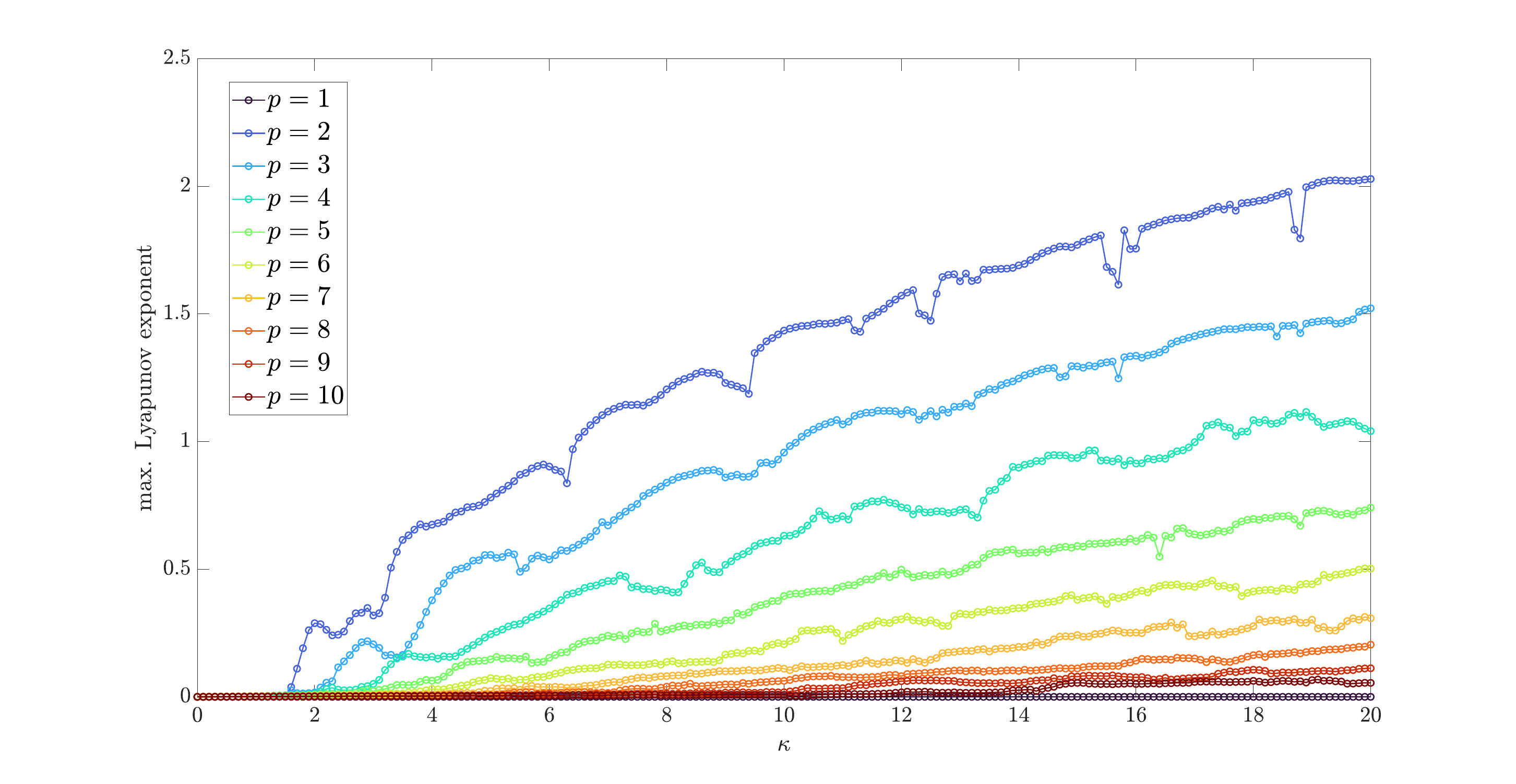}
        \caption{Classical Lyapunov exponent for different integer values of $p$ from 1 to 10.
        The $p=1$ curve lies along the $LE=0$ axis.}
        \label{fig:max lyapunov exp. for mod p-spin with 1.0 to 10 with 1 div model with 10000 kick and 200 div upto k 20}
    \end{subfigure}
    \hfill
    \begin{subfigure}[H]{0.45\textwidth}
        \centering
       \includegraphics[trim={4cm 0 0 0},clip,width=9.5cm,height=5cm]{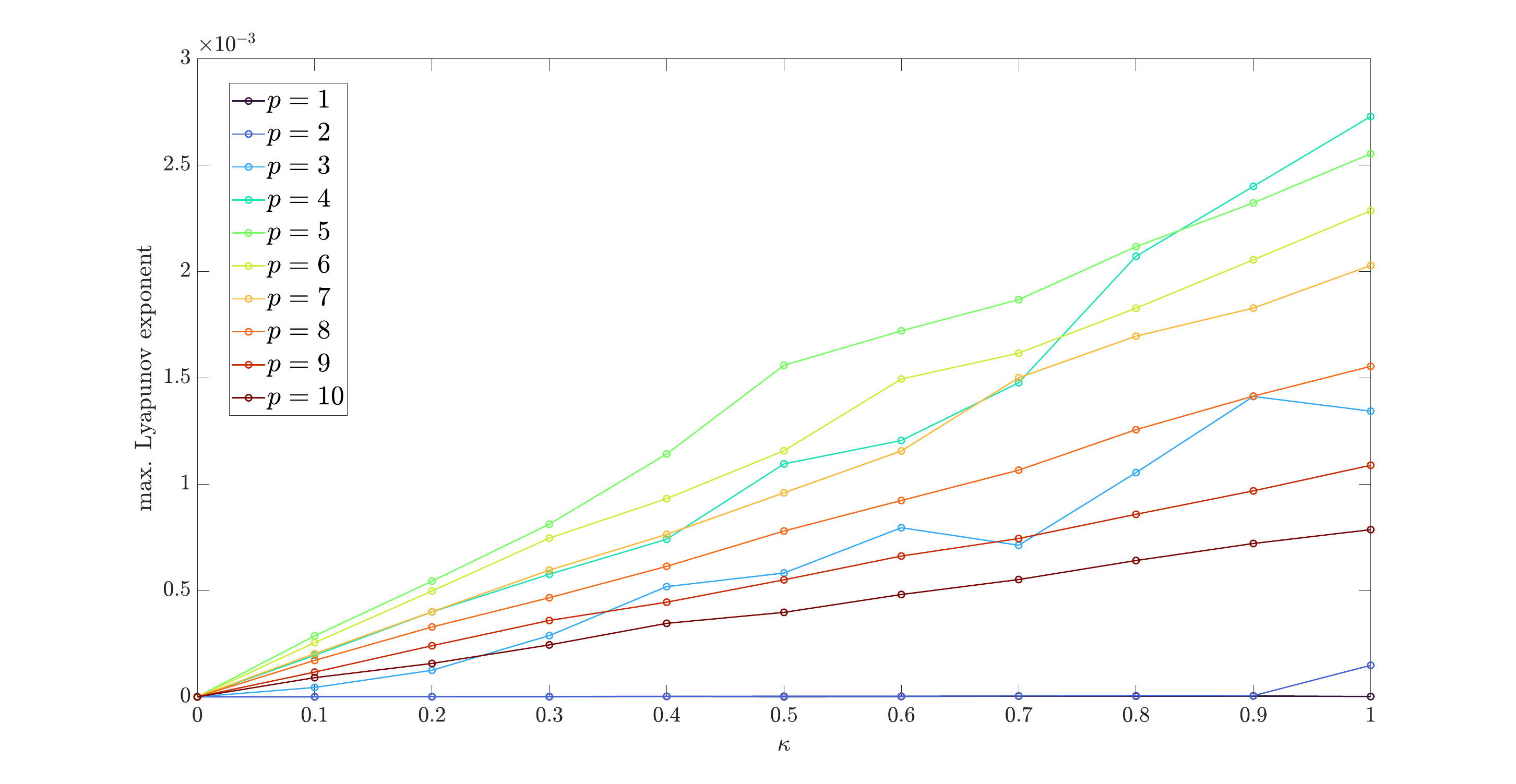}
        \caption{Classical Lyapunov exponent for different integer values of $p$ from 1 to 10.}
      \label{fig:max lyapunov exp. for mod p-spin with 1.0 to 10 with 1 div model with 10000 kick and 200 div upto k 1}
    \end{subfigure}
    \caption{Classical Lyapunov exponent for different values of $p$ and $\alpha=\pi/2$. For a given $\kappa$ and $p$, 289 uniformly distributed initial points on the phase space  were each evolved for $10^4$ kicks. The maximum Lyapunov exponent was calculated by taking the average over the whole sphere.}
    \label{fig:overall}
\end{figure*}

Lyapunov exponents (LEs) characterize the rate of separation of infinitesimally close classical trajectories \cite{strogatz_2019_nonlinear}. In a Hamiltonian system, these exponents quantify the rate of divergence or convergence of nearby trajectories within the system's phase space, offering   insights into its long-term behavior. Since  Hamiltonian systems are area preserving maps, a positive Lyapunov exponent signifies chaos. Conversely, negative exponents signify stability, with trajectories converging towards invariant tori or other regular structures in phase space.  

The tangent map corresponding to $F$  in  \eqref{eq:classical_eqn} is given by 
\begin{equation}\label{eq:tangent map}
    \delta\vec{X}_{n+1} = M(\vec{X}_n) \delta \vec{X}_n,
\end{equation}
where
\begin{equation}\label{eq:jacobian matrix}
    M(\vec{X}_n) = \Bigg(\frac{\partial\vec{X}_{n+1}}{\partial \vec{X}_n}\Bigg).
\end{equation}
The explicit form of a tangent map for $\alpha=\pi/2 $ takes the form 
\begin{align}\label{eq:tangent map for pi/2}
\begin{split}  
M(\vec{X}_n) = \hspace{6cm}\\
\begin{pmatrix}
-Z_n \,  C_1\sin(C_2) + Y_n \,  C_1 \cos(C_2) & \sin(C_2) & \sin(C_2) \\
-Y_n \, C_1 \sin(C_2) - Z_n \,  C_1 \cos(C_2) & \cos(C_2) & -\sin(C_2) \\
-1 & 0 & 0
\end{pmatrix},
\end{split}
\end{align}
where $C_1= \kappa (p-1)|X_n|^{p-2}$ and $C_2 = \kappa X_n|X_n|^{p-2}$. 
The largest LE is given by 
\begin{equation}\label{eq:LE}
    \lambda(\kappa,\alpha) = \ln \Bigg[\lim_{N\to\infty} | x_+(N)|^{1/N}\Bigg],
\end{equation}
where $x_+(N)$ is the largest eigenvalue of the matrix product $\prod_{n=1}^{N} M(\vec{X}_n)$. To obtain the LE from our numerical simulations, we set $N=10^4$. A positive LE characterizes chaotic dynamics. For strongly chaotic orbits, one can further simplify   \eqref{eq:LE}  \cite{Constantoudis_and_Theodorakopoulos_1997}.

\section{Numerical simulations }\label{section: numerical investigations}

Our main objective is to explore the critical degree of non-linearity necessary for the  system \eqref{eq:modulus_hamiltonian} described by the equations of motion \eqref{eq:classical_eqn}
to exhibit chaotic behavior.
Throughout our study we  consistently set $\alpha=\pi/2$, while varying $\kappa$ across various sets of initial conditions.
We use \eqref{eq:LE} to compute the largest Lyapunov exponent via numerical simulation.  Our approach involved the initialization of 289 states uniformly distributed across the phase space. Each such point  is evolved for $10^4$ kicks with the  LE subsequently  calculated. The global Lyapunov exponent was then determined by averaging across the entirety of the phase space. A stroboscopic picture of results across the range of $p$-values we studied is shown in
figure~\ref{fig:classical phase space structure}.

\subsection*{Case 1 : $1\leq p \leq2$}\label{section:p less than 2}

The form of the Hamiltonian \eqref{eq:modulus_hamiltonian} allows us to consider fractional values of  $p$.  
We considered 200 distinct values of $\kappa$ ranging from 0 to 20. The system loses its integrability for any non-zero value of $\kappa$. Our analysis indicates that for $p=1$, despite being non-linear,  the system fails to exhibit chaos:  its largest Lyapunov exponent remains 0 for all values of $\kappa$. This phenomenon is particularly noteworthy when $p=1$ and $\alpha=\pi/2$, for which the argument of trigonometric function in  \eqref{eq:classical_eqn}, $\kappa\zeta_n |\zeta_n|^{p-2} \to \kappa\frac{X_n }{|X_n|}$. Consequently the non-linear term flips sign  contingent upon the top's orientation along the x-direction, resulting in instantaneous switching of the sign in the dynamical equations \eqref{eq:classical_eqn}. Therefore, the kicked part of the Hamiltonian is only able to perform instantaneous switching but not a twist in the traditional sense. 
While this action imparts instantaneous switching, the lack of twisting or mixing, crucial for inducing chaos in Hamiltonian systems, inhibits the system from transitioning into chaotic dynamics. For this case, we also do not have fixed   and period-4 points, as mentioned in section  \ref{section:p-kicked top model}. These points are absent for this case as $\kappa\frac{X_n }{|X_n|}$ is not defined for $X=0$. This situation stands in notable contrast to three-body one-dimensional self-gravitating systems \cite{LEHTIHET198693,Burnell:2002ps}, which exhibit a mild form of chaos for a Hamiltonian whose interaction experiences a similar sign flip as particles interchange positions.

The behaviour of the $p=1$ case can also be understood from  a series expansion of \eqref{eq:classical_eqn} in powers of $\zeta_n^{p-1}$:
  \begin{eqnarray}\label{eq:series expansion of classical equations }
   \nonumber X_{n+1} & \approx & \mu_n - Y_n\chi - \frac{1}{2}\mu_n\chi^{2} + \frac{1}{6}Y_n\chi^{3} + \mathcal{O}(\chi^{4(p-1)}), \\
  \nonumber Y_{n+1} & \approx & Y_n - \mu_n\chi - \frac{1}{2}Y_n\chi^{2} + \frac{1}{6}\mu_n\chi^{3} + \mathcal{O}(\chi^{4(p-1)}), \\
  Z_{n+1} & = & \zeta_n. 
  \end{eqnarray} 
where $\chi=\kappa(\zeta)^{p-1}$. For $p=1$, the dependence on the non-linear term vanishes, since   $\zeta_n^{p-1} = 1$. The switching of the sign comes from the orientation of $\zeta_n$ ( $X_n$ for $\alpha=\pi/2$).  The upper left panel \ref{fig:p1.0} of 
figure~\ref{fig:classical phase space structure} shows a stroboscopic map of the trajectories for $p=1$. Despite the complex structure of the global phase space (see Fig. \ref{fig:p1.0}), the LE for this case is zero. This case is further discussed in section \ref{section:fractal-like structures}. The non chaotic behaviour for $p=1$ is explicitly shown in figure~\ref{fig:max lyapunov exp. for mod p-spin with 1.0 to 2.0 with 0.1 div model with 10000 kick and 200 div upto k 20}, which depicts the variation in the largest Lyapunov exponent as a function of $\kappa$ for different degrees of nonlinearity, $1\leq p \leq 2$. 
 
For small values of $\kappa$ the system exhibits  chaos for all $p$ values except for $p=1$ and $p=2$. The $p=1$ case remains regular for all values of $\kappa$. For   $p=2$ (the standard kicked top \cite{haake1987}) global chaos remains absent for $\kappa < 2.2$.   The small-$\kappa$  behaviour is shown in figure~\ref{fig:max lyapunov exp. for mod p-spin with 1.0 to 2 with 1 div model with 10000 kick and 200 div upto k 1}. Except for $p=1,2$, we obtain positive values of largest LE for all $\kappa>0$. The LE for $p=1.9$ lies very close to the $LE=0$ axis, but is nonzero for all values of $\kappa$.
It is also interesting to note that, $p=2$ demonstrates saturation of the largest LE for $\kappa \geq 2$. 
These results suggest that even for the near-integrable (small $\kappa$) case  the system exhibits chaotic behaviour.

\subsection*{Case 2 : $ p \geq 2$}\label{section:p greater than 2}

Non-linearity stands as a fundamental prerequisite for a system to exhibit chaos. Contrary to the previous case, for $p>2$, as the degree of non-linearity increases, the chaotic domain on the phase space becomes confined to an increasingly narrower band. This can be seen in the global phase space structure of the classical kicked top in panels \ref{fig:p4}, \ref{fig:p7}, and \ref{fig:p10} of figure~\ref{fig:classical phase space structure}. These figures reveal a reduction in the chaotic dynamics  with increasing $p$. 

The global phase space structure is constructed using 289 initial points each evolved for 400 kicks (see Fig \ref{fig:classical phase space structure}). To quantify  global chaos we used  a numerical technique similar to the previous case. We initialize 289 states and evolved for $10^4$ kicks and calculate the largest LE using Eq. \eqref{eq:LE}.

In figure~\ref{fig:max lyapunov exp. for mod p-spin with 1.0 to 10 with 1 div model with 10000 kick and 200 div upto k 20} we plot   the largest LE against $\kappa$
for integer values of $p$ ranging from 1 to 10. For $\kappa>2$, the largest LE decreases with increasing   $p$.
For small $\kappa < 2$ the situation is notably different: the largest LE is for $p=5$, as seen in figure~\ref{fig:max lyapunov exp. for mod p-spin with 1.0 to 10 with 1 div model with 10000 kick and 200 div upto k 1}.

This interesting behaviour can be understood from inspection of the series expansion
in \eqref{eq:series expansion of classical equations }. The system's non-linearity stems from the dependence of the system on   $\zeta_n^{(p-1)}$. For $\alpha=\pi/2$, $\zeta_n$ is simply $X_n$. As $p$ increases, $X_n^{(p-1)} (\in [0,1])$ diminishes, reducing the dependence of the non-linear component of the $n^{th}$ step on the $(n+1)^{th} $ step. Consequently, the system's chaotic behaviour decreases with increasing $p,$ ultimately transitioning towards a purely regular oscillating system in the limit   $p\to \infty$. For $\zeta_n < 0$, we obtain a similar expansion but with the opposite sign for odd powers. 

A parallel analysis reveals that for sufficiently large $\kappa$ the maximum influence of the non-linear term is manifest for $p=2$, such that $\zeta_n^{(p-1)}$ reaches its peak value. Contrastingly, for any other values of $p ( > 2)$, the dependence on non-linear term is reduced. This observation underscores the fact that the kicked top system attains its highest degree of chaos when $p=2$, corresponding to the original kicked top model. In this scenario as well, the system exhibits chaotic behavior for any $\kappa>0$. However, the extent of global chaos observed in the system is notably reduced compared to the preceding case of $1\leq p \leq 2$ for the same $\kappa$ value.

\section{Fractal-like structures}\label{section:fractal-like structures}

\begin{figure*}[htbp] 
    \centering
    \begin{subfigure}[H]{0.48\textwidth}
        \centering
        \includegraphics[width=1.02\columnwidth,trim={1.55cm 0 1.5cm 1.2cm},clip]{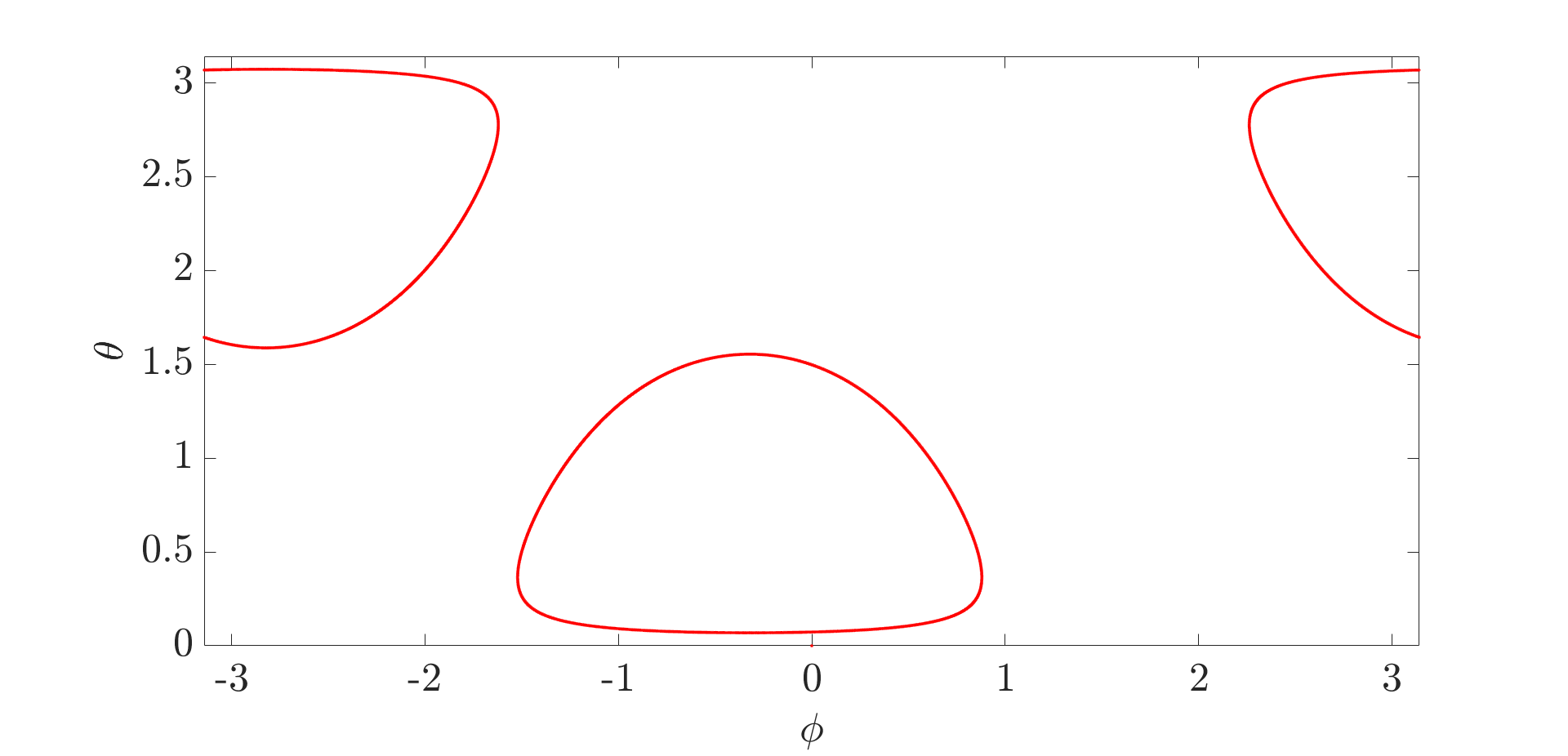}
        \caption{$(\theta,\phi)= (0.5,-1.5)$}
        \label{Appendix:fig 1 fractal-like}
    \end{subfigure}
    \hfill
    \begin{subfigure}[H]{0.48\textwidth}
        \centering
        \includegraphics[width=1.02\columnwidth,trim={1.55cm 0 1.5cm 1.2cm},clip]{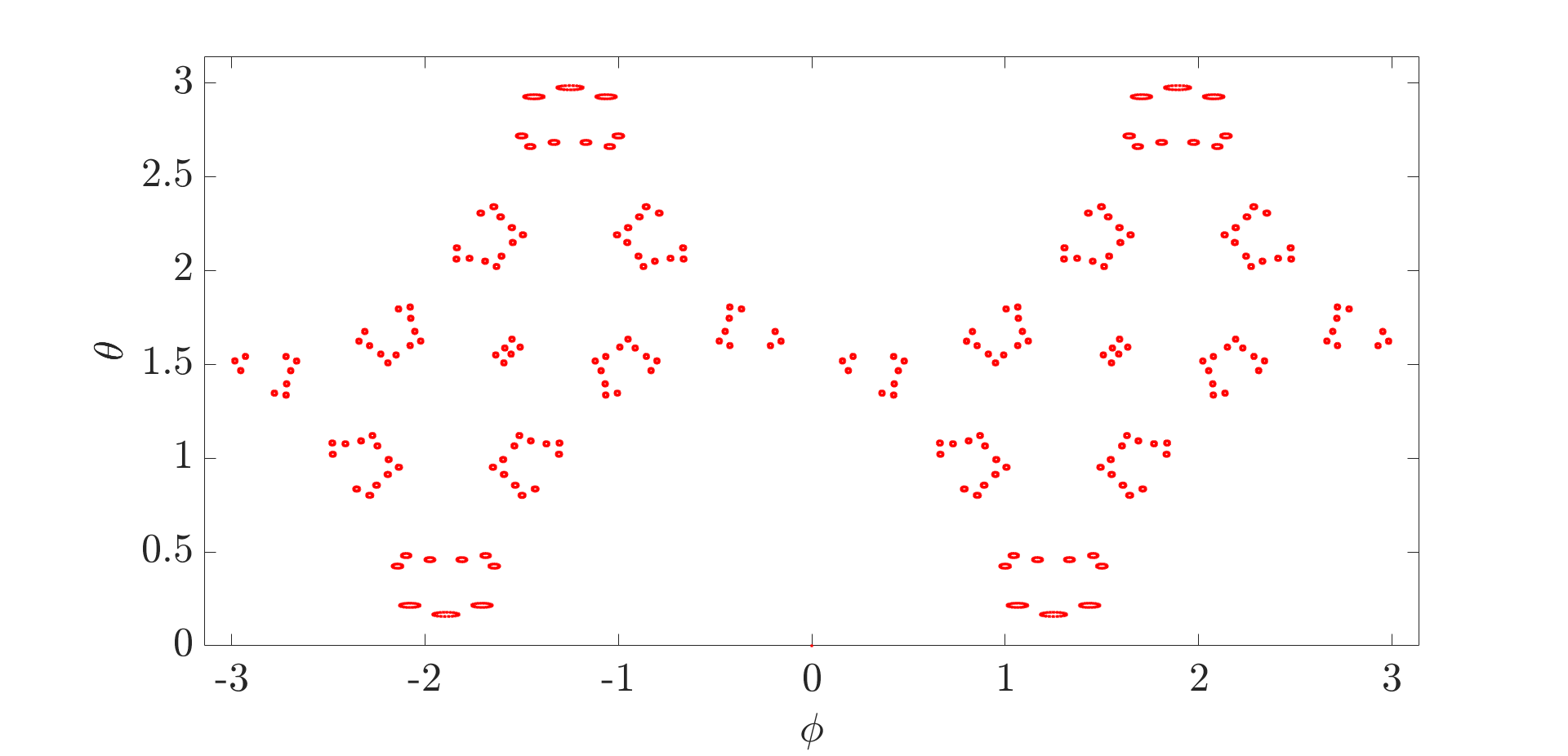}
        \caption{$(\theta,\phi)= (2.2,-1)$}
        \label{Appendix: fig 2 fractal-like}
    \end{subfigure}
    \\
    \vspace{5mm}
    \begin{subfigure}[H]{0.48\textwidth}
        \centering
        \includegraphics[width=1.02\columnwidth,trim={1.55cm 0 1.5cm 1.2cm},clip]{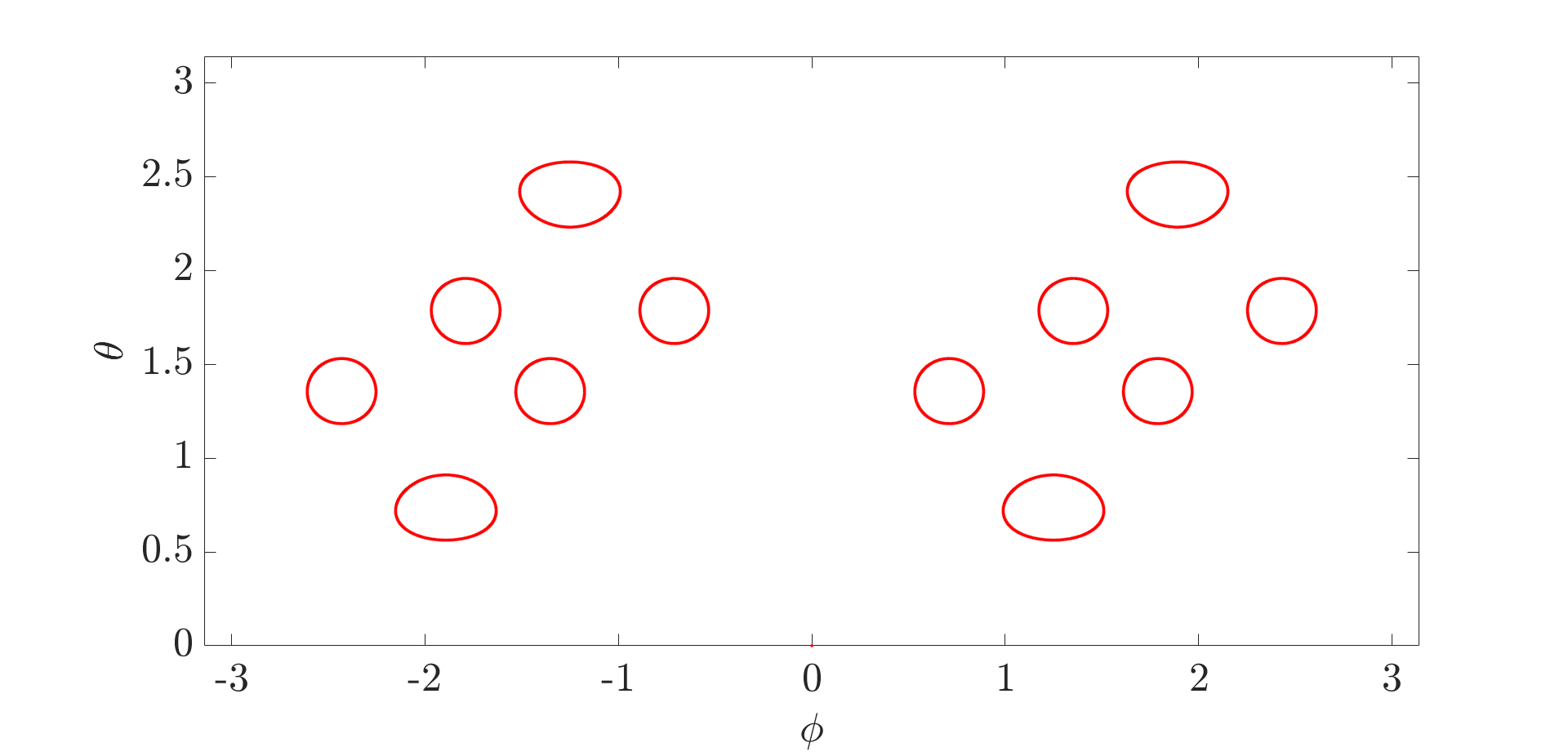}
        \caption{$(\theta,\phi)= (2.25,2.0)$}
        \label{Appendix: fig3 fractal-like}
    \end{subfigure}
    \hfill
    \begin{subfigure}[H]{0.48\textwidth}
        \centering
        \includegraphics[width=1.02\columnwidth,trim={1.55cm 0 1.5cm 1.2cm},clip]{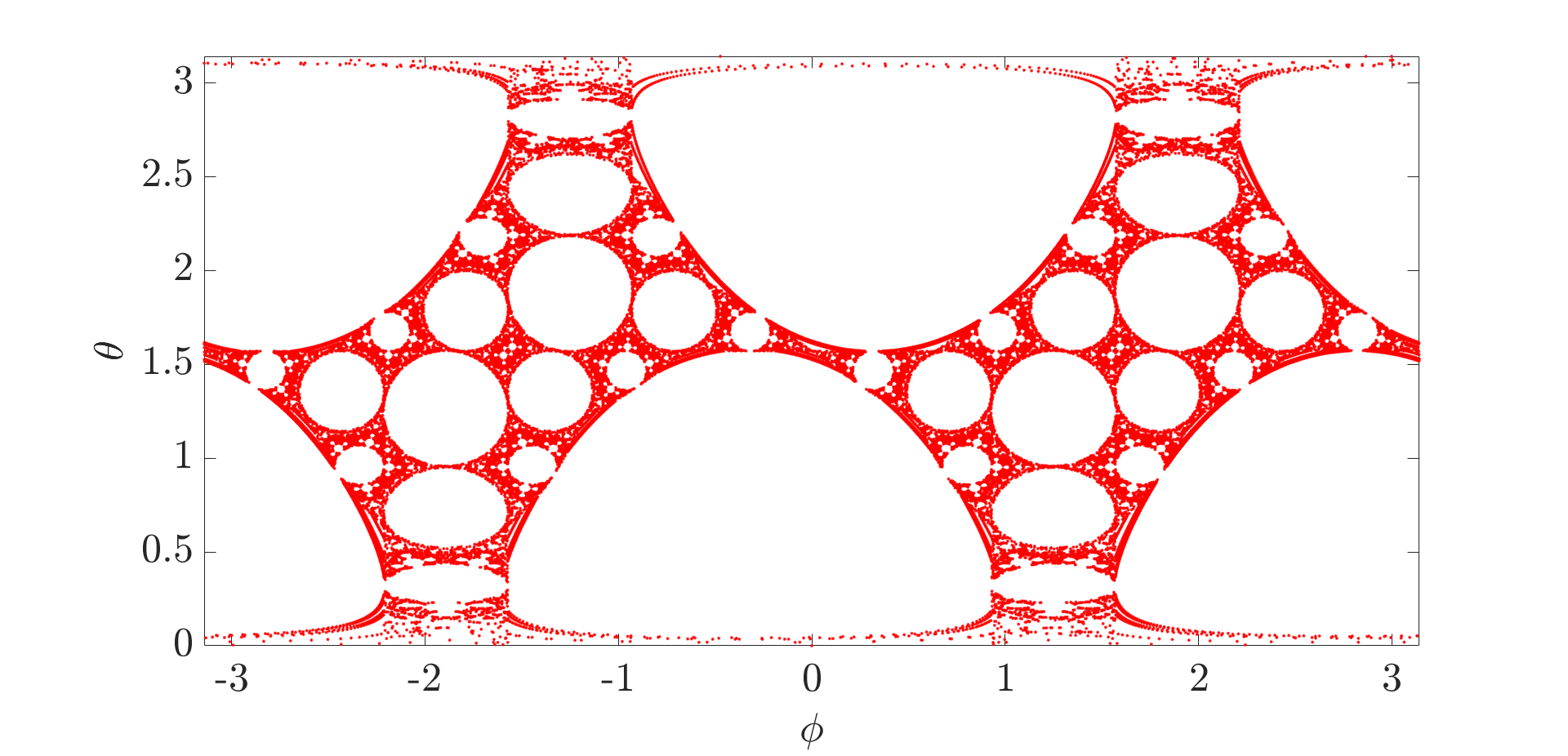}
        \caption{$(\theta,\phi)= (\pi/2-\delta,-\delta)$}
        \label{fig:fractal-like}
    \end{subfigure}
    \caption{Classical stroboscopic map for $p=1$, $\kappa=2.5$ and different initial points.  Each initial point was evolved for 80000 kicks. In (d), $\delta$ is taken $10^{-4}$ where effects of $X=0$ plane (discontinuity) can be seen on the dynamics on the phase space leading to the complex structure. See text for more discussion.}\label{appendix fig: classical dynmaics for p=1and kappa=2.5}
    \label{fig:overall}
\end{figure*}
The  $p=1 $ case presents a particularly intriguing scenario. Despite its non-linear nature, as discussed in Section \ref{section:p less than 2}, the system does not exhibit chaotic behavior. Instead, its dynamics display fractal-like patterns. However, precisely defining these patterns within conservative Hamiltonian systems is challenging due to the absence of an attractor. This case was previously studied in the context of understanding circle mappings on the spherical phase space \cite{scott_hamiltonian_2001,scott_hamiltonian_2003}. To illustrate the complex structures, we present the long-term dynamics of various initial points for 
$\kappa=2.5$ in Figure \ref{appendix fig: classical dynmaics for p=1and kappa=2.5}. In certain regions of the overall structure, such as the one highlighted in Figure \ref{fig:fractal-like} around the coordinates $(\theta,\phi = \pi/2,0)$, we observe repeated structures. This is because, for $X=0$, the limit does not exist for the system. Therefore, the behavior of initial points close to the $X=0$ plane differs due to the presence of this discontinuity.

Previous studies have aimed to determine whether fractals can emerge within conservative Hamiltonian systems. One such example is the piece-wise linear map known as the sawtooth standard map \cite{Chen_meiss_1989,chen_meiss_1990,ashwin_97}. The non-linearity stems from the discontinuity along the dominant symmetry line, $x=0$. The sawtooth map on the cylinder $[0,1) \times \mathbb{R}$ is defined by  
\begin{eqnarray}\label{eq:sawtooth map}
   \nonumber p_{n+1}&=&p_n + F(x_n) \\  
              x_{n+1}&=& x_n + p_{n+1} \,\, \text{mod} 1
\end{eqnarray}
where
\begin{equation}\label{eq:F(x) of sawtooth map}
    F(x) = \begin{cases}
    k(x-\frac{1}{2}) &  \text{even integer} \\
    0 & x=0,1
\end{cases}.
\end{equation}
and $k$ is a positive parameter representing the kick strength. The function $F(x) $ is discontinuous at $x=0$, a dominant line.  
The sawtooth model consists of elliptic islands around every periodic orbit if it avoids encountering the discontinuity line $x=0$ \cite{ashwin_97}. However, if an orbit encounters this discontinuity, it receives a jump start. This discontinuity leads to sensitive dependence within the system and one can get complex structures similar to the model studied here. Nonetheless, for the sawtooth map as well, the Lyapunov exponent remains zero.

Another study by Giorgilli \cite{giorgilli_relevance_1988}  examines the $\mu$-dependent family of mappings of the torus $\bold{T^2}$ into itself. The mapping is given by
\begin{align}
\left.
\begin{array}{rl}
x' &= x + y \\
y'  &=  y - \mu \sin(2 \pi x')
\end{array}
\right\} & \pmod{1}
\end{align}
where $\mu$ is a real parameter. Numerical investigations indicated  non-integral dimensions of these mappings for $\mu=0.12$, with the initial points lying close to the unstable point $(0.5,0)$ \cite{giorgilli_relevance_1988}. The box-counting method was used to calculate the Hausdorff dimension of the system. However, theoretically, the Hausdorff dimension of chaotic orbits in conservative systems should be an integral number \cite{ott_1981}. This inconsistency between theoretical and numerical predictions was explained as an effect of finite observation of scale and time.

\section{Discussion}\label{section: discussion}
According to   Kolmogorov-Arnold-Moser (KAM) theory\cite{kolmogorov_1954,arnold_proof_1963,moser_1962}, any non-linear perturbation of an integrable system breaks its integrability. Non-integrability, while being  a minimum requirement,  does not guarantee chaotic behavior. We therefore explored the minimal level of non-linearity requisite for a system to exhibit chaos.
We have presented a modified version of the well-established kicked top and its chaotic behaviour, and obtained the critical exponent necessary to induce chaotic dynamics for a broad range of parameters.   

Our  generalization of the kicked top exhibited new behaviour dependent on the exponent $p$ of the non-linear part of the Hamiltonian \eqref{eq:modulus_hamiltonian}. An investigation 
via a stroboscopic map indicated that the fixed points and period-4 points were same for all the values of $p$ except for $p=1$. 

A noteworthy aspect of this model pertains to the $p=1$ case. In this scenario, the system becomes inherently nonlinear compared to the original kicked top, leading to a loss of integrability for any positive values of $\kappa$. We explain this behaviour for $\alpha=\pi/2$, where perturbation effectively acts as switching instead of twist. Despite the absence of chaotic dynamics in this setting, the $p=1$ top exhibits complex behavior characterized by fractal-like structures.
We demonstrated the sensitivity of select initial states on the sphere in section \ref{section:fractal-like structures}, 
highlighting the nuanced dynamics of the system. For any $p$ value greater than $1$ and $\alpha=\pi/2$, the model exhibits chaotic behavior, with degree of chaos increasing as $p$ rises until $p=2$. Another observation is that the modified kicked top has a positive Lyapunov exponent for all positive values of $\kappa$ within the range $1<p<2$, whereas for $p=2$, chaos is not manifest until $\kappa\approx2.2$.

We recover the original kicked top for $p=2$. However, for $p>2$, the top exhibits complex behavior.For smaller $\kappa$, as we increase the value of $p$, the system directly transits into chaotic dynamics for all $p$, with the exception of $p=2$. This behavior mirrors that observed within the  range $1 < p \leq 2$. Nevertheless, for $p>2$, the degree of chaos is notably reduced compared to the aforementioned scenario. With increasing $p$, the global chaos in the system diminishes, ultimately leading to a regular coupled oscillator in the limit  $p\to \infty$. This phenomenon arises because, with increasing $p$, the impact of the perturbation decreases, as evidenced by linearizing the dynamical map in \eqref{eq:series expansion of classical equations }. This characteristic is unique to the Hamiltonian, as the dynamical variables ($\vec{X}$, $\vec{Y}$, and $\vec{Z}$) are constrained within the range of [0,1]. 

We have also discussed the impact of the discontinuity for the $p=1$ case and its effect on the phase space structure. Our analysis indicates that any system with instantaneous switching can create a fractal-like structure on the phase space. 

Whether the discrepancy between theoretical and numerical studies observed in the past persists in this model is an intriguing question for future research. 
Investigating whether the modified top exhibits similar dependence on the exponent   $p$ in the quantum realm would also be particularly interesting. For even integer values of $p$, our model converges to the kicked $p$-spin model, where investigations were conducted for $p=2,3,4$ \cite{Poggi_chaos_2021}. Given the importance of the $p$-spin model in the field of quantum information \cite{jorg_Pujos_2010,Poggi_simulation_2021} and quantum annealing \cite{Daniel_2017}, in conjunction with the richness of the dynamics of the present model, it is natural to explore the modified quantum kicked top with modulus non-linear exponent.

\section*{Acknowledgements}
This work was supported in part by the Natural Sciences and Engineering Research Council of Canada (NSERC). Wilfrid Laurier University and the University of Waterloo are located in the traditional territory of the Neutral, Anishnawbe and Haudenosaunee peoples. We thank them for allowing us to conduct this research on their land. A.A. would like to express sincere gratitude to Prof. James Meiss for his invaluable assistance and insightful feedback, which contributed to this research.

\section*{Code and Data availability}
 The code and data that support this work are available on request. Please contact corresponding author for code and data.

\section*{Competing interests}
The authors declare no competing interests.

\vspace{5mm}

\appendix
\section*{Appendix : Classical equation of motion}\label{Appendix: Classical equation of motion}
\setcounter{equation}{0} 
\renewcommand{\theequation}{A \arabic{equation}} 

The classical Hamiltonian for the modified kicked top is 
\begin{equation}\label{eq:classical_modulus_hamiltonian}
H = \frac{\alpha J_{y}}{\tau} + \frac{\kappa {|J_{z}|}^p}{pj^{p-1}} \sum_{n=-\infty}^{\infty} \delta (t-n\tau) .
 \end{equation}
By expressing the classical angular momentum  $\vec{J}$ as the cross product 
$\vec{J}=\vec{r}\cross \vec{p}$ of   position ($\vec{r}$) and momentum $\vec{p}$ we get 
\begin{equation}\label{eq:classical_modulus_hamiltonian_subsituation}
H = \frac{\alpha}{\tau} (zp_x - xp_z)+\frac{\kappa}{p j^{p-1}} {|(xp_y - yp_z)|}^p\sum_{n=-\infty}^{\infty} \delta (t-n\tau) 
 \end{equation}
\vspace{2mm}
Using Hamiltonian's equations of motion, $\Dot{q_i}=\partial H/ \partial p_i$ and $\Dot{p_i}=\partial H/ \partial q_i$, 
\begin{subequations}
\begin{align}
\dot{x} &= \frac{\alpha}{\tau}z -\frac{\kappa}{j^{p-1}}{|J_{z}|}^{p-2}J_z y  \sum_{n=-\infty}^{n=\infty} \delta (t-n\tau) \\ 
\dot{p_x} &= \frac{\alpha}{\tau} p_z -\frac{\kappa}{j^{p-1}}{|J_{z}|}^{p-2}J_z P_y \sum_{n=-\infty}^{n=\infty} \delta (t-n\tau)  \\
\dot{y} &= \frac{\kappa}{j^{p-1}}{|J_{z}|}^{p-2}J_z x \sum_{n=-\infty}^{n=\infty} \delta (t-n\tau) \\  
\dot{p_y} &= \frac{\kappa}{j^{p-1}}{|J_{z}|}^{p-2}J_z P_x   \sum_{n=-\infty}^{n=\infty} \delta (t-n\tau)  \\
\dot{z} &= \frac{\alpha}{\tau}x \\
\dot{p_z} &= \frac{\alpha}{\tau}p_x
\end{align}\label{apendix eq: time derivative of position and momentum vector}
\end{subequations}
Using the above equations we find 
\begin{subequations}
\begin{align}
\Dot{J_x}  &= \frac{\alpha}{\tau}J_z -\frac{\kappa}{j^{p-1}}{|J_{z}|}^{p-2}J_z J_y \sum_{n=-\infty}^{n=\infty} \delta (t-n\tau) \\ 
\Dot{J_y}  &= \frac{\kappa}{j^{p-1}}{|J_{z}|}^{p-2}J_z J_x \sum_{n=-\infty}^{n=\infty} \delta (t-n\tau) \\
\Dot{J_z}  &= - \frac{\alpha}{\tau}J_x
\end{align}\label{apendix eq: time derivative of angular momentum}
\end{subequations}
for  the classical equations of motion of the angular momentum vector. 
To obtain the classical equations for the normalized angular momenta $\vec{J}_i/j$, we divide each of the preceding equations by $j$, to get
\begin{subequations}
\begin{align}
\Dot{X}  &= \frac{\alpha}{\tau}Z -\kappa{|Z|}^{p-2}ZY \sum_{n=-\infty}^{n=\infty} \delta (t-n\tau) \\ 
\Dot{Y}  &= \kappa{|Z|}^{p-2}ZX \sum_{n=-\infty}^{n=\infty} \delta (t-n\tau) \\
\Dot{Z}  &= - \frac{\alpha}{\tau}X
\end{align}\label{apendix eq: time derivative of normalized angular momentum}
\end{subequations}
where   $ \vec{X}_i = \vec{J}_i/j $.

To compute the stroboscopic map with initial conditions $[X_n,Y_n,Z_n]$, we integrate Eq.\eqref{apendix eq: time derivative of normalized angular momentum} for one time period $\tau$, in the interval of $[\tau,\tau-\epsilon]$, followed by $[\tau-\epsilon,\tau + \epsilon]$ and then taking the limit $\epsilon \to 0$.
In the interval of $[\tau,\tau-\epsilon]$ Eq.\eqref{apendix eq: time derivative of normalized angular momentum} reduces to 
\begin{subequations}
\begin{align}
\Dot{X}  &= \frac{\alpha}{\tau}Z \\
\Dot{Y}  &= 0\\
\Dot{Z}  &= - \frac{\alpha}{\tau}X
\end{align}\label{apendix eq: time derivative of normalized angular momentum for tau-epsilon time}
\end{subequations}
Integrating this with   $[X_n,Y_n,Z_n]$ as the initial state yields $[\Tilde{X}_n,\Tilde{Y}_n,\Tilde{Z}_n]$ where 
\begin{subequations}
\begin{align}
\Tilde{X}_n  &= X_n\cos{\alpha} + Z_n\sin{\alpha}\\
\Tilde{Y}_n  &= Y_n\\
\Tilde{Z}_n  &= Z_n\cos{\alpha} - X_n\sin{\alpha}
\end{align}\label{apendix eq: state after integration for tau-epsilon time}
\end{subequations}
In the subsequent time interval $[\tau-\epsilon,\tau + \epsilon]$, we integrate the kicked part of    \eqref{apendix eq: time derivative of normalized angular momentum} 
\begin{subequations}
\begin{align}
\Dot{X}  &=  -\kappa{|Z|}^{p-2}ZY \sum_{n=-\infty}^{n=\infty} \delta (t-n\tau) \\ 
\Dot{Y}  &= \kappa{|Z|}^{p-2}ZX \sum_{n=-\infty}^{n=\infty} \delta (t-n\tau) \\
\Dot{Z}  &= 0
\end{align}\label{apendix eq: time derivative of normalized angular momentum in epsilon time}
\end{subequations}
with initial conditions $[\Tilde{X}_n,\Tilde{Y}_n,\Tilde{Z}_n]$. 
The solution of \eqref{apendix eq: time derivative of normalized angular momentum in epsilon time} is given by 
\begin{eqnarray}\label{apendix eq: state after integration for epsilon time}
\nonumber  X(t)  &= & \Tilde{X}_n \cos(\Omega) - \Tilde{Y}_n \sin(\Omega)\\ 
\nonumber Y(t)  &= &\Tilde{Y}_n \cos(\Omega) + \Tilde{X}_n \sin(\Omega)\\
Z(t)  &= & \Tilde{Z}_n 
\end{eqnarray}
where $\Omega=\kappa |\Tilde{Z}|^{p-2}\Tilde{Z}f_H(t-\tau)$ and $f_H(t-\tau)$ is a Heaviside function with value $0$ for $t<\tau$, and 1 when $t>\tau$. Writing $\zeta_n = Z_n \cos(\alpha) - X_n \sin(\alpha)$ and $\mu_n = X_n \cos(\alpha) + Z_n \sin(\alpha)$, substituting $[\Tilde{X}_n,\Tilde{Y}_n,\Tilde{Z}_n]$ from   \eqref{apendix eq: state after integration for tau-epsilon time} into  \eqref{apendix eq: state after integration for epsilon time}
and taking the limit $\epsilon \to 0$ we get  
 \begin{subequations}
\begin{eqnarray}
\nonumber X_{n+1}&=&\mu_n\cos\bigg(\kappa \zeta_n |\zeta_n|^{p-2} \bigg) - Y_n\sin\bigg(\kappa \zeta_n |\zeta_n|^{p-2}\bigg), \\  
\nonumber Y_{n+1}&=&Y_n\cos\bigg(\kappa \zeta_n |\zeta_n|^{p-2} \bigg) + \mu_n\sin\bigg(\kappa \zeta_n |\zeta_n|^{p-2}\bigg), \\
 Z_{n+1}&=& Z_n \cos(\alpha) - X_n \sin(\alpha)
\end{eqnarray}\label{apendix eq: classical_eqn}
\end{subequations}   
for the final stroboscopic map.

  \bibliographystyle{elsarticle-num} 
  \bibliography{ref}






\end{document}